# Design and implementation of an islanded hybrid microgrid system for a large resort center for Penang Island with the proper application of excess energy

SK. A. Shezan[1] | Rawdah[1] | S. Shafin Ali[1] | Ziaur Rahman[2]

[1]School of Engineering, RMIT University, Melbourne, Victoria, Australia

[2]School of Science, RMIT University, Melbourne, Victoria, Australia

**Correspondence**
SK. A. Shezan, School of Engineering, RMIT University, Melbourne, VIC, Australia.

**Funding information**
Federal Government of Australia

**Abstract**

The energy demand is growing daily at an accelerated pace due to the internationalization and development of civilization. Yet proper economic utilization of additional energy generated by the Islanded Hybrid Microgrid System (IHMS) that was not consumed by the load is a major global challenge. To resolve the above-stated summons, this research focuses on a multi-optimal combination of IHMS for the Penang Hill Resort located on Penang Island, Malaysia, with effective use of redundant energy. To avail this excess energy efficiently, an electrical heater along with a storage tank has been designed concerning diversion load having proper energy management. Furthermore, the system design has adopted the HOMER Pro software for profitable and practical analysis. Alongside, MATLAB Simulink had stabilized the whole system by representing the values of 2068 and 19,072 kW that have been determined as the approximated peak and average load per day for the resort. Moreover, the optimized IHMS is comprehended of Photovoltaic (PV) cells, Diesel Generator, Wind Turbine, Battery, and Converter. Adjacent to this, the optimized system ensued in having a Net Present Cost (NPC) of $21.66 million, Renewable Fraction (RF) of 27.8%, Cost of Energy (COE) of $0.165/kWh, $CO_2$ of 1,735,836 kg/year, and excess energy of 517.29MWh per annum. Since the diesel generator lead system was included in the scheme, a COE of $0.217/kWh, $CO_2$ of 5,124,879 kg/year, and NPC of $23.25 million were attained. The amount of excess energy is effectively utilized with an electrical heater as a diversion load.

## 1 | INTRODUCTION

Energy demand is expanding altogether because of the fast development and progression of industrialization and urbanization. The ordinary petroleum product based force stations emanates a lot of greenhouse gasses (GHG), which are a danger to the climate and humanity. Along these lines, for clear reasons, there is a move of enthusiasm toward different non-customary and sustainable power assets like the breeze, sun based, wave, biomass, geothermal, and hydro energy. Advances identified with wind power and sun oriented forces are more full-grown and promptly accessible among environmentally friendly power assets and the portion of these assets are expanding in frameworks and microgrids all around the globe. As per the Renewable Global Status report, 33.1% of worldwide power is enhanced from sustainable power assets while the current 67.2% of power is gotten from atomic force plants and petroleum derivatives.[1] Around 40% of the populace daily routines without power and the larger part experience in country regions. Despite the methodologies executed by numerous individuals of the nations in Sub-Saharan Africa and South Asia, there are still around 2 billion residents without energy availability in those areas.[2] Be that as it may, individuals living in far off islands of Malaysia depend on diesel generators for power due to lacking responsiveness to the public framework coming about because of the massive foundation tax of the transmission lines.[3] Close by, the traveler layer in the islands of Malaysia depends vigorously on diesel generators for power gracefully for the day.[4] Regardless, the high activity and support cost of diesel plants, unnecessary $CO_2$ exudation and the temperamental market cost of diesel fuel incite the framework to be ecologically perilous and extravagant.[5,6] In comparison, the diesel price in the Malaysian islands is nearly double the initial price relative to that of the city center.[7] Thus, an IHMS can perform a vital role in supplying available power service to the tourist zone on the relative Malaysian islands. The proper utilization of excess energy generated by IHMS has recently been considered a major concern in the design of IHMS.[8,9] For this purpose, many energy system designs find a hybrid power system in which two or more renewable energy sources are paired along with battery storage systems and often diesel generators as backups.[10-12] However, the excess energy will damage the storage battery due to overloading, which can also affect the devices attached to the device if not unloaded. To maintain a balance of power in the system, the over-supplied energy should be removed or used for other practical purposes.[13,14] A wind-diesel-battery system in an off-grid configuration for two residential hotels in Cameron highland, Malaysia has a value of 8.7 kW as the peak load, as reported by Shezan.[15] The system resulted in 33.6% unused excess energy. Demiroren and Yilmaz have conducted research to supply electricity to Gokceada Island, Turkey and their optimized wind turbine system generated 74.3% excess electricity of the entire electrical production.[16] The authors suggested a wind turbine system with a grid-connected configuration to handle the energy. In addition, a hybrid system comprising of PV-diesel-battery was suggested by Ismail, M.S., M. Moghavvemi, and T.M.I. Mahlia,[17] for supplying 77.45 kWh per day to an isolated house in Langkawi Island in Malaysia and resulted in 31.33% excess electricity per year of total electricity production, which was simply dumped. However, the technological viability of an IHMS, grid-connected IHMS, and off-grid system for large hotels located in the sub-tropical coastal area of Queensland has been studied solely considering the NPC and payback period only instead of concentrating on the utilization excess energy.[18,19]

A PV-wind-diesel-battery system considering seven renewable scenarios was investigated by Mustafizur in a North American off-grid community.[20,21] After the study, the authors obtained 67.8%, 28.6%, 32.3%, 19.7%, 16.7%, 4.3%, and 11.6% surplus energy of total generation when the renewable resources were 100%, 80%, 65%, 50%, 35%, 21%, and 0%, respectively. A summary of the proportion of excess electricity in the form of percentage for numerous hybrid system configurations analyzed from the diverse inspection is illustrated in Table 1.

Therefore, with a potent usage of surfeit energy through diversion load, this study suggested an IHMS with an off-grid design for Penang Hill resort, Malaysia having a load requirement of 19,072 kWh/day.

Henceforth, the recent survey focuses on the following research tags:

i. In consideration of the corresponding Payback Period (PBP) and NPC, the suitable RES components for the proposed scheme need to be chosen.[21]

**TABLE 1** A summary of the proportion of surfeit electricity in the form of percentage for numerous hybrid system designs conducted earlier

| Category of the hybrid system | Percentage of surfeit energy (%) | Remarks | Reference |
| --- | --- | --- | --- |
| Wind-diesel-battery | 31.2 | Grid-connected | 4 |
| Stand-alone wind | 67.3 | Suggested grid | 15 |
| PV-diesel-battery | 13.56 | Grid-connected | 16 |
| PV-diesel-battery | 42.7 | Dumped | 19 |
| PV-diesel-battery | 23.5 | Unutilized | 19 |
| PV-wind | 74.9[a] | Dumped | 20 |
| PV-wind-diesel | 30.7[b] | Dumped | 20 |
| PV-wind-diesel | 37.3[c] | Unutilized | 20 |
| Wind-diesel | 19.8[d] | Dumped | 20 |
| PV-wind-diesel | 16.4[e] | Grid-connected | 20 |
| PV-wind-hydro | 71[f] | Dumped | 21 |
| PV-wind-hydro-diesel | 17.9[g] | Grid-connected | 21 |

[a]100% renewable scenario.
[b]80% renewable scenario.
[c]65% renewable scenario.
[d]50% renewable scenario.
[e]35% renewable scenario.
[f]69% PV penetration.
[g]51% PV penetration.

ii. The most cost-effective feasible technology for RES needs to be predicted and then compared with other similar technologies. In this case, the comparison should be made taking the $CO_2$ emissions, COE, and NPC into account for various RES.[23]

iii. Acknowledging competent energy supply and appropriate regulation of RES, the feasibility of the proposed system must be examined.

## 2 | SITE DESCRIPTION AND RESOURCE ASSESSMENT

### 2.1 | Penang Hill Resort

For this respective research, a substantial bungalow styled resort and hotel, Penang Hill resort, located in Penang Island has been elected. The geographical location of the resort is 5°24′52.2″N 100°19′45.1″E, as shown in Figure 1. Penang Island, is the main constituent island of the Malaysian state of Penang. Located at the Malacca Strait, off the northwestern coast of Peninsular Malaysia, it is separated from the mainland by the Penang Strait. The island is home to nearly half of Penang's population; the city of George Town, which covers the island and the five outlying islets, is Malaysia's third-largest city by population.[24,25]

### 2.2 | Estimation of load profile

Installation of diesel generators providing electricity throughout the respective location resulted in the unavailability of the hourly load data at the resort. Thus, the resort's hourly load profile was determined taking into account the number of rooms, the size of the rooms, the type of electrical appliances used, the type of accommodation, and the variation in tourist presence during different seasons. The resort is a chalet-style structure with 312 rooms and suites, two restaurants (Sri Nelayan and Natahari), three bars (beach bar and sunken pool bar), one ballroom (Shahzad all, I and II), which can accommodate approximately 500 guests, and six other separate meeting rooms (Mukut, I and II and Penang, I and II). All the 312 rooms are facilitated with an air-cooling system, Television, hairdryer, coffee maker, tube light, table lamp, and telephone. Each ballroom and meeting rooms have air conditioning, overhead projector, slide projector, microphone, tube light, compact fluorescent lamp (CFL) bulb, desktop computer, and laptop computer. The electrical appliances available in the bar and restaurants are air conditioning, freezer, refrigerator, tube light, coffee maker, and CFL bulb. Air conditioning, laptop and desktop computer, printer and scanner, toast maker, electric oven, and fridge are fitted in the resort administrate office.[26] Furthermore, approximated peak capacitive load and average load per day were observed to be 2068 kW and 19,072 kW reciprocally. Table 2 represents a range of electrical devices employed along with the power rating while Figure 2 illustrates the estimated annual load profile data.

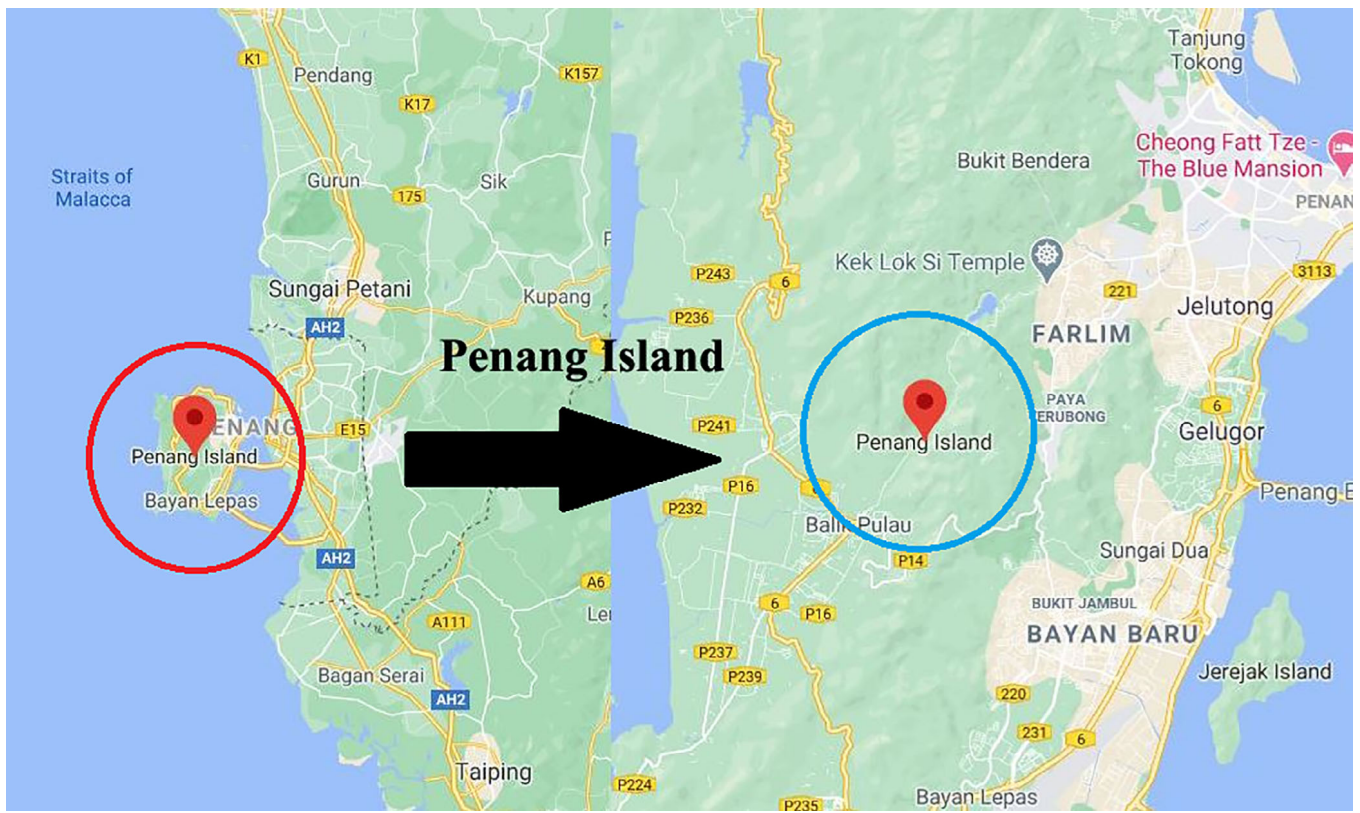

**FIGURE 1** The geographical position of Penang Island, Malaysia
[Source: Edited version using reference from Google map] [Color figure can be viewed at wileyonlinelibrary.com]

**TABLE 2** Estimated load along with the rated power[24]

| Electrical appliances | No. of components | Running hour/day | Rated power output (Watt) | Electrical Appliances | No. of components | Running hour/day | Rated power (Watt) |
| --- | --- | --- | --- | --- | --- | --- | --- |
| Air conditioner | 400 | 15 | 1400 | Hair straightener | 312 | 5 | 2500 |
| Television | 282 | 7 | 175 | Coffee maker | 312 | 5 | 1500 |
| Light | 1000 | 17 | 30 | Fan | 600 | 10 | 80 |
| Lamp | 620 | 8 | 20 | Microwave Oven | 7 | 7 | 3000 |
| Bulb | 60 | 10 | 45 | PC | 6 | 15 | 120 |
| Stereopticon | 3 | 6 | 250 | Laptop | 5 | 15 | 80 |
| Slither projector | 3 | 6 | 320 | Versatile printing machine | 5 | 5 | 100 |
| Mike | 7 | 4 | 15 | Refrigerator | 320 | 15 | 450 |
| Refrigerator | 320 | 15 | 450 | Versatile printing machine | 5 | 5 | 100 |

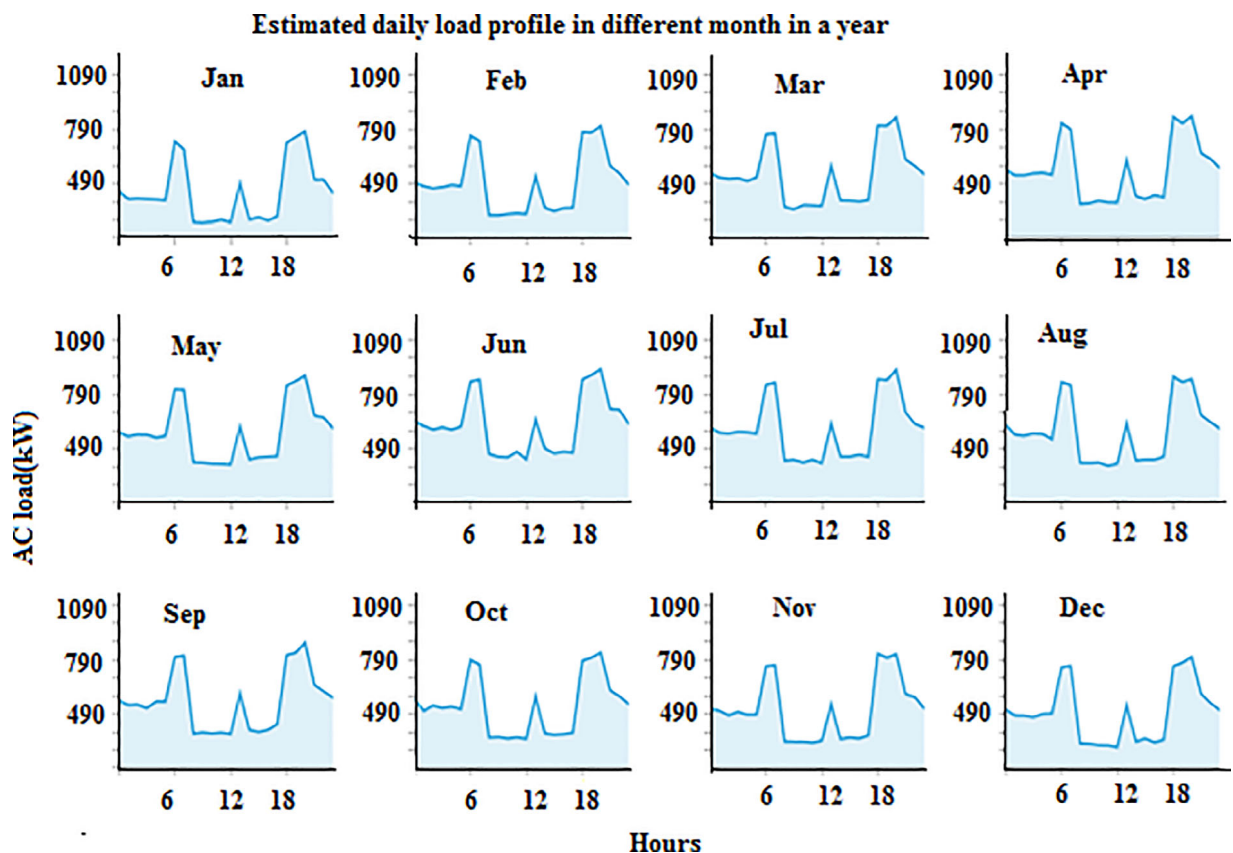

**FIGURE 2** Approximated monthly load profile data daily for the resort [Color figure can be viewed at wileyonlinelibrary.com]

## 2.3 | Solar energy

In terms of generating an efficient solar energy conversion and calculation of production capacity for the system, Solar radiation data analysis is necessary.[27] Moreover, for this research, daily average global solar irradiance for the period 2006–2018 has been assembled from the Malaysian Meteorological Department (MMD) at Hospital Kota Tinggi since there are no astrological whereabouts in Penang reef. The average solar radiation for this station is determined to be 4.20kWh/m$^2$/day annually. Daily clearness index and average solar radiation per month for the relative island are demonstrated in Figure 3. From the meteorological department, the annual average solar radiation data can be collected. In addition to this, the maximum power from a solar panel can be calculated using the following equation[28]:

$$P_{mp} = \eta_{PV} G_\beta A \quad (1)$$

where the surface area of the PV module is denoted by A. $P_{mp}$ refers to the maximum power from a solar panel. $\eta_{PV}$ terms to the efficiency of the silicon-based PV cell. $G_\beta$ defines the global horizontal solar irradiation.

$$\eta_{PV} = \frac{V_{OC} I_{SC} FF}{P_{in}} \quad (2)$$

where $V_{OC}$ is the open-circuit voltage. $I_{SC}$ is the short-circuit current. FF is the fill factor. $P_{in}$ is input power.

## 2.4 | Wind energy

Wind energy act as free sustainable energy resources with no air or water pollution. Evaluation of wind speed profile for the desired wind farm location is essential before its construction.[29,30] Since there is no climatological location in Penang landmass, so the wind speed data from 11 years (2006–2014) has been obtained from MMD at Mersinias.

## 2.5 | Dissemination of wind speed frequency

The Wind Speed Frequency Distribution (WSFD) remains to approach the economic viability and wind energy potentiality of any specific area. To apt, the wind profile for a range of data, different WSFD representations such as the Weibull, the Rayleigh, and the Lognormal are extensively used.[21,31-33] The Weibull distribution function is widely used for wind speed analysis because of its great flexibility besides simplicity.[34] Intended for this study, the two parameters of Weibull distribution stay chosen. The possibility of wind at a given velocity V is determined via the probability distribution function (PDF) and PDF is expressed by the following impression.[24]

$$f(V) = \left(\frac{k}{c}\right)\left(\frac{V}{c}\right)^{k-1} \exp\left[-\left(\frac{V}{c}\right)^k\right] \quad (3)$$

The possibility that the wind velocity is identical to or lesser than V or inside a given wind speed assortment is indicated by Cumulative Distribution Function (CDF) of wind speed V and that the CDF is expressed by following impression.[19,21]

$$F(V) = 1 - \exp\left[-\left(\frac{V}{c}\right)^k\right] \quad (4)$$

Subsequently, within these findings, the shape and scale factors were reckoned by the standard deviation technique using Equations (4) and (5) from and are expressed below.[34]

$$k = \left(\frac{\sigma}{V_m}\right)^{-1.086} \quad (5)$$

Following this, obtained values of standard deviation, Weibull shape, and scale factor, furthermore practicable wind speed and wind speed carrying maximum energy and wind power density are shown in Table 3. Figure 4 shows the PDF plot of wind speed for ages between 2004 and 2018.[28]

## 2.6 | Techno-economic analysis

To recommend the maximum amalgamation of apparatuses in the IHMS, techno-economic investigation is very essential. Therefore, HOMER Pro software has been utilized to achieve this motive, and is further dependent on Net Present Cost (NPC) given by,[31]

$$C_{npc,tot} = \frac{C_{ann,tot}}{CRF(i, R_{proj})} \quad (6)$$

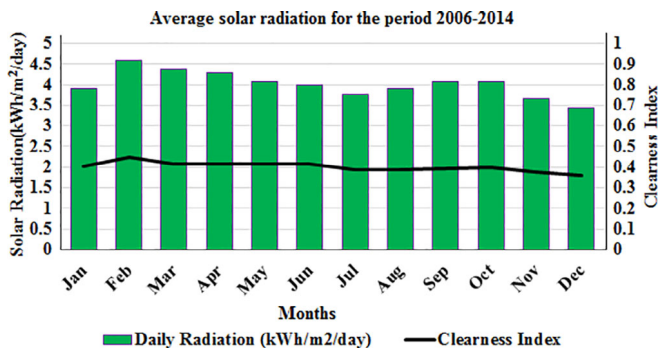

**FIGURE 3** Typical clearness index and daily solar radiation per month at Penang Island [Color figure can be viewed at wileyonlinelibrary.com]

**TABLE 3** Weibull parameters along with wind power concentration for Penang Island

| Months | 1 | 2 | 3 | 4 | 5 | 6 | 7 | 8 | 9 | 10 | 11 | 12 |
|---|---|---|---|---|---|---|---|---|---|---|---|---|
| Σ | 2.56 | 2.35 | 2.32 | 1.56 | 1.54 | 1.55 | 1.63 | 1.67 | 1.56 | 1.50 | 1.76 | 2.57 |
| K | 3.80 | 4.93 | 3.04 | 706 | 6.13 | 6.49 | 5.48 | 5.67 | 6.33 | 6.68 | 4.45 | 3.91 |
| c (m/s) | 6.87 | 6.25 | 4.13 | 4.16 | 3.07 | 5.27 | 3.29 | 4.58 | 3.28 | 3.06 | 6.26 | 4.70 |
| $V_{mp}$ (m/s) | 5.41 | 4.76 | 3.62 | 3.07 | 2.98 | 6.18 | 7.17 | 3.46 | 3.20 | 5.98 | 3.08 | 6.07 |
| $V_{max,E}$ (m/s) | 7.56 | 5.83 | 4.87 | 3.31 | 3.21 | 3.41 | 3.48 | 3.76 | 4.42 | 3.45 | 3.54 | 5.63 |
| P/A (W/m$^2$) | 111.6 | 85.9 | 40.9 | 16.4 | 14.9 | 18.2 | 18.6 | 23.9 | 18.3 | 14.9 | 18.3 | 55.1 |

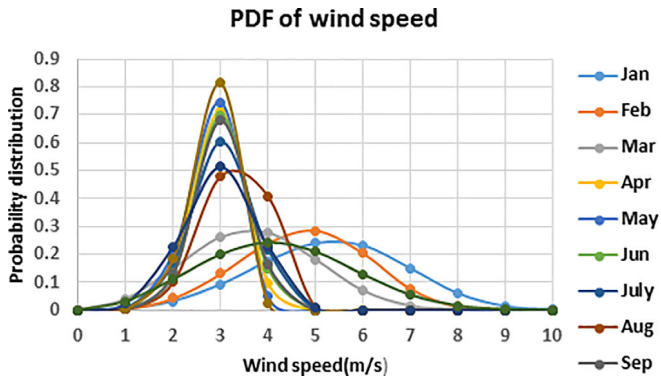

**FIGURE 4** The plot for Probability Distribution Function (PDF) of wind speed data in years between 2004 and 2018 [Color figure can be viewed at wileyonlinelibrary.com]

$C_{ann,tot}$ from the equation is the total annual cost ($/year), $i$ is the annual real interest rate (%), $R_{proj}$ is the project lifetime (year), CRF represents capital recovery factor and is denoted by the following equation.[31]

$$CRF(i,N) = \frac{i(1+i)^N}{(1+i)^N - 1} \qquad (7)$$

The number of years and yearly real interest rate is denoted by $N$ and $i$ from the above expression, respectively. To further include, the minimum interest rate in the data processing is not adopted by HOMER Pro instead definite interest rate is calculated from the actual significance interest rate using the subsequent equation as shown in Equation (8).

$$i = \frac{i' - f}{1 + f} \qquad (8)$$

Furthermore, from the above Equation (8), the 12-month price increasing rate and the insignificant interest rate is expressed as $f$ and $i'$. The cost of energy (COE) defines the price per kWh of electricity.[21] And hence, isolating the total annualized cost by every year electricity served to load measures the COE, which remains equated as follows

$$COE = \frac{C_{ann,tot}}{L_{ann.load}} \qquad (9)$$

where $C_{ann,tot}$ → Total annual cost ($/year). $L_{ann.load}$ → Total annual load (kWh).

The following subsidiary equations have been familiarized to estimate the $CO_2$ emissions from the hybrid energy system as:

$$tCO_2 = 3.667 \times m_f \times HV_f \times CEF_f \times X_c \qquad (10)$$

where $tCO_2$ equals the amount of $CO_2$ emissions, $m_f$ defines fuel quantity (Liter), $HV_f$ refers to fuel heating value (MJ/L), $CEF_f$ is the carbon emission factor (ton carbon/TJ), and $X_c$ stands to be oxidized carbon fraction. Another factor must be also be contemplated that in 3.667 g of $CO_2$ 1 g of carbon is encompassed.

## 2.7 | The typical strategy of HOMER Pro and it is component specifications

Thereby, with battery, diesel generator, ac load profile solar, wind, and converter, IHMS has been primarily configured as shown in Figure 5. This is used to reduce the additional transportation and labor cost; the worth of diesel price for Malaysian Island is deemed to be advanced than the actual price of the mainland.[7]

## 2.8 | Photovoltaic system

The PV size estimated for this study ranges between 0 and 1000 kW with a step size of 50 kW. The temperature co-efficient, derating factor, operating temperature considered for the PV panel were −0.5/°C, 80%, and 47°C, respectively. Table 4 demonstrates the further detail of the PV system with the technical specifications and benchmark of the specifications. Table 4. displays details of the price of the PV system published in Reference 28 that was used in simulations.

## 2.9 | Battery storages

For the simulation of the hybrid system, Trojan IND 17 batteries were used. The number of strings considered was ranging from 0 to 10 with each parallel string consisting of 40 batteries. The lifetime of each battery was 9300 kWh. Table 4 demonstrates the further detail of the battery with the technical specifications and benchmark of the specifications. The battery efficiency is 95%, which denotes that maximum energy can be extracted from battery storage under the stochastic conditions. The battery charge efficiency is equal to the square root of the battery round-trip efficiency, hence:

$$\eta_{batt,c} = \sqrt{\eta_{batt,rt}} \qquad (11)$$

where $\eta_{batt,c}$ is battery charge efficiency. $\eta_{batt,rt}$ is battery round-trip efficiency.

The battery discharge efficiency is equal to the square root of the storage round-trip efficiency, hence:

$$\eta_{batt,d} = \sqrt{\eta_{batt,rt}} \qquad (12)$$

where $\eta_{batt,d}$ is battery discharge efficiency. $\eta_{batt,rt}$ is battery round-trip efficiency.

## 2.10 | Wind turbine

Emergya Wind Technologies (EWT) Direct Wind 52/54 wind turbine was considered in the simulations. The nominal production power of the turbine is 250 kW through rated wind speed 8 m/s and cut-in wind speed 3.5 m/s. Furthermore, the number of wind turbines considered for simulation varies between 0 and 8. Besides, the pivot altitude and lifetime for the turbine were found as 50 m and 20 years, respectively. A brief representation of the detailed cost of the wind turbine with the benchmark of the specification is shown in Table 4.

**FIGURE 5** Proposed design model for IHMS with diversion load control option

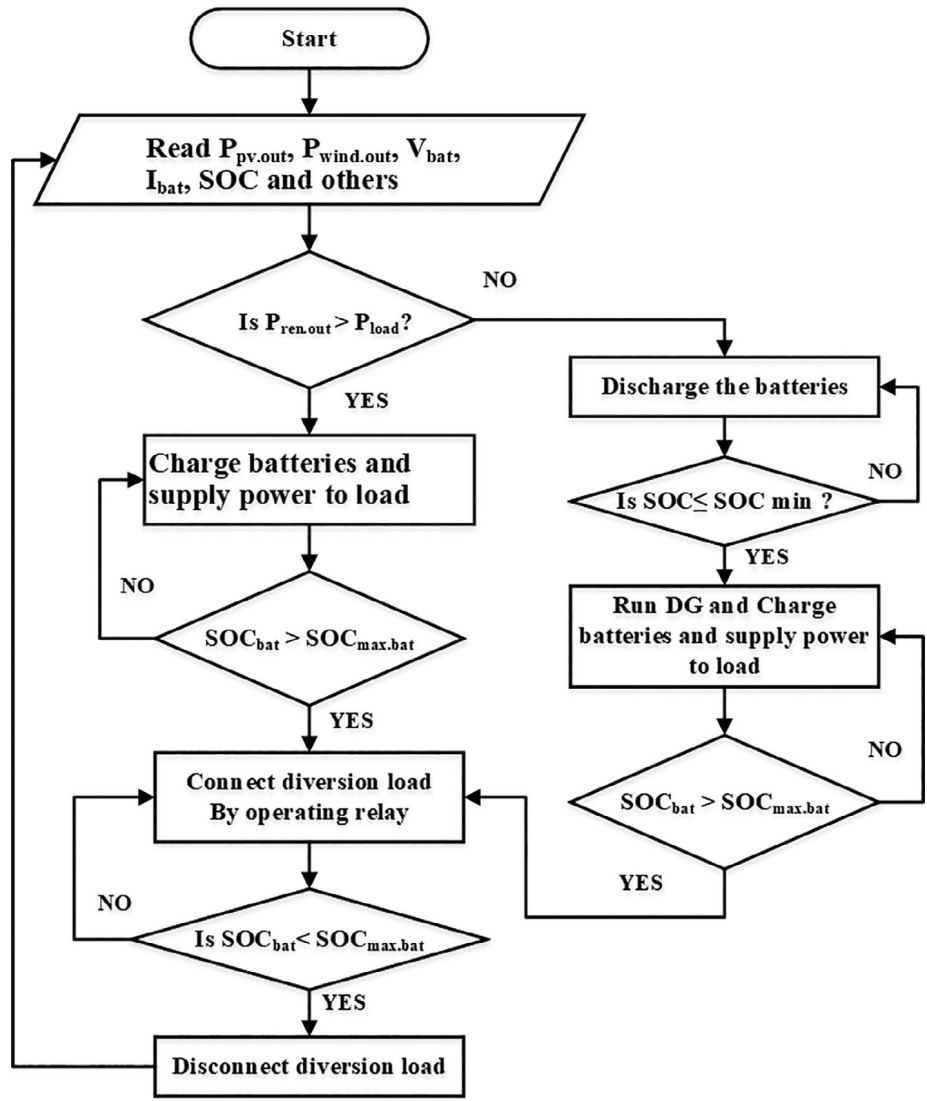

## 2.11 | Diesel generator

In this article, three diesel generators having a rating 0–500 kW have been considered in the HOMER Pro model with a step size of 50 kW. With the size of generators, the location of the project, and manufacturers, the primary capital and additional cost of generators vary significantly. Moreover, minimal running hours and lowest load ratio of each generator were assumed to be 15,000 h and 25%, respectively. Table 4 provides detailed information about diesel generators with the benchmark of the specification.

## 2.12 | Bidirectional converter

The size of the bidirectional converter was within a rating of 0–1000 kW through a phase size of 100 kW. Respectively, $890/kW and $800/kW was taken to be the primary capital and extra cost of the converter. The lifetime and efficiency of the converter were considered 15 years and 95%, respectively. Table 4 provides detailed information about the converter with the benchmark of the specification.

## 2.13 | System control, economics, and constraints

Commercial depreciation, diesel off operation, and cycle charging, are the control parameters, which consents several generator functionalities and system with two types of wind turbines. Subsequently, 30% and 50% outputs referring to that of solar and wind, respectively, have been considered throughout the research. In addition, a zero-penalty cost was taken for carbon dioxide.

**TABLE 4** Economic and technical specifications of the components

| PV [54] | | |
|---|---|---|
| Description | Specifications | Benchmark [29] |
| Capital cost | $2000/kW | $1500–2500/kW |
| Replacement cost | $2000/kW | $1500–2500/kW |
| O & M cost | $10/kW/year | $8–12/kW/year |
| **Batteries [2]** | | |
| Description | Specifications | Benchmark [30] |
| Nominal voltage | 6 Volt | 6–12 Volt |
| Capacity | 1231 Ah | 1000–1300 Ah |
| Capital cost | $1200/unit | $1000–1500/unit |
| Replacement cost | $1170/unit | $1000–1500/unit |
| O & M cost | $10/year | $10–15/year |
| Efficiency | 96% | 91%–96% |
| **Wind Turbine [55, 56]** | | |
| Description | Specifications | Benchmark [31] |
| Capital cost | $375,000/unit | $365,000–400,000/unit. |
| Replacement cost | $262,500/unit | $250,000–350,000/unit |
| O & M cost | $7500/year | $7000–8500/year |
| Cut-in wind speed | 3.5 m/s | 3–3.5 m/s |
| Nominal wind speed | 8.4 m/s | 7.5–8.7 m/s |
| Cut out wind speed | 18 m/s | 14–19 m/s |
| **Diesel Generator [57]** | | |
| Description | Specifications | Benchmark [32] |
| Brand name | Cummins | |
| Size | 0–500 kW | 0–500 kW |
| Capital cost | $220/kW | $200–320/kW |
| Replacement cost | $200/kW | $180–300/kW |
| O & M cost | $0.03/h | $0.02–0.05/h |
| **Converter [58]** | | |
| Description | Specifications | Benchmark [33] |
| Capital cost | $890/kW | $850–950/kW |
| Replacement cost | $800/kW | $750–850/kW |
| O & M cost | $10/kW/year | $8–12/kW/year |
| Efficiency | 97% | 95%–97% |

## 2.14 | Effective utilization of excess renewable energy

The excess energy generated from IHMS can be effectively utilized with diversion load and proper energy management.[9,13] When the batteries are fully charged, the charge control of a PV array is fairly genuine where the controller disconnects the PV array. Thus, the batteries are protected from being overcharged and being damaged. For limiting the load within the generator to regulate the charge rate similarly, chargers and inverters are connected to the generator power. A turbine will "freewheel" with an increase in its rpm (revolutions per minute) and voltage for an unloaded condition and hence, the turbine output must be connected to the load always without being persistent in allowing the overcharging of the batteries. In addition, a diversion load control can be effectively used to accomplish this purpose more precisely.[8]

## 2.15 | Anticipated design system for the consumption of superfluous energy by the diversion load

To avail of a surplus amount of electricity from renewable energy sources, an electrical heater provided with a storage tank as diversion load is one of the methods. For this reason, diversion load control must be incorporated with the other equipment in the IHMS.[13] Figure 5 illustrates the model diagram of the projected stand-alone IHMS incorporated with diversion load control. Furthermore, the policy of the off-grid IHMS with diversion load comprising of control and power management is described in the flow chart in Figure 6. With this, the power output of diesel generators, wind, PV, and State of Charge (SOC) for batteries is always monitored by the master controller following the respective block diagram and power management strategy. However, the master regulator permits the battery to store sustainable energy when the power output from renewable sources ($P_{ren.out}$) is superior to that of load demand ($P_{load}$). On the other hand, the master regulator tends to avail storage batteries as fall back when the power output from sustainable resources ($P_{ren.out}$) is a reduced amount of load demand ($P_{load}$) until the relative storage batteries have reached minimum SOC. Hence or otherwise, the diesel generator will come into operation for the convenience of delivering the mandatory load demand.

## 3 | RESULT AND DISCUSSION

According to the inputs being sorted from the optimization results starting from the lowest NPC to highest NPC, HOMER Pro gives a simulation of all the possible combinations. The obtained optimization results are represented in Tables 5 and 6 having the categorized form, respectively. In contrast to this, the surpass optimized IHMS consists of 200 batteries, three diesel generators having a rating of 100, 300, and 500 kW, a 600 kW PV, four wind turbines, and a converter of 500 kW. These optimization results take the NPC of $17.55 m, RF of 31.5%, COE of $0.165/kWh, EE of 517.29MWh (11.3% of total electrical production), and emits yearly $CO_2$ of 1,735,836 kg. The second-best optimization, the wind-DG-batt system has a renewable fraction of 28.3% and NPC approximating the best optimization result, yet six wind turbines contribute only 6.32% of total electrical production. In comparison to this, the third-best optimization entails 700 kW photovoltaic cell, two 500 kW diesel generator, and five wind turbines with a renewable fraction of 34.9%, the NPC and COE are more advanced than the superlative case. Notably, $CO_2$ emission, according to this case, is the lowest comparing to another optimal hybrid system.

**FIGURE 6** A complete flowchart representing power administration for IHMS with diversion load

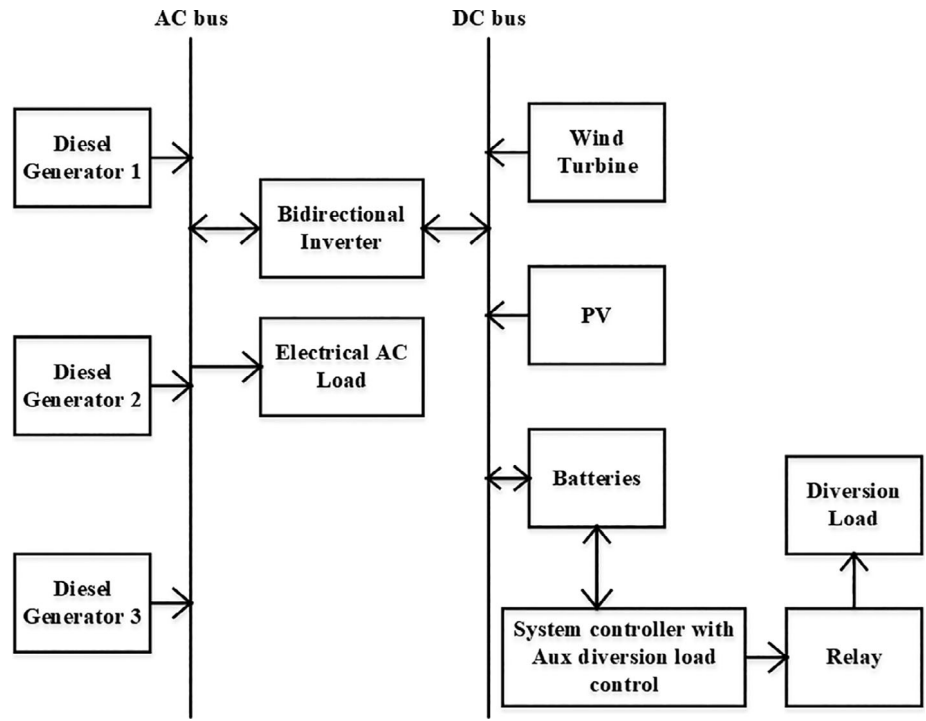

**TABLE 5** The adjusted hybrid sustainable system fixed by NPC

| Position | PV rating (kW) | Wind (No) | GEN1 (kW) | GEN2 (kW) | GEN3 (kW) | Batt | Conv (Kw) | Dispatch | COE ($/kWh) | NPC ($) | RF (%) | EE (MWh/year) |
|---|---|---|---|---|---|---|---|---|---|---|---|---|
| 1 | 600 | 4 | 100 | 300 | 500 | 200 | 500 | Combined dispatch | 0.165 | 19,547,522 | 33.5 | 517.19 (7.8%) |
| 2 | 0 | 6 | 100 | 300 | 500 | 200 | 500 | Cycle charging | 0.170 | 19,857,376 | 29.3 | 779.35 (12.1%) |
| 3 | 700 | 5 | 0 | 500 | 500 | 280 | 600 | Lod following | 0.180 | 20,493,432 | 37.9 | 757.22 (11.5%) |
| 4 | 1000 | 0 | 100 | 300 | 500 | 240 | 400 | Generator order | 0.185 | 20,768,800 | 18.9 | 82.62 (1.5%) |
| 5 | 0 | 6 | 0 | 500 | 500 | 280 | 600 | Combined dispatch | 0.190 | 20,809,708 | 26.6 | 776.41 (12%) |
| 6 | 400 | 4 | 500 | 300 | 300 | 0 | 400 | Cycle charging | 0.195 | 20,921,564 | 25.2 | 623.40 (9.6%) |
| 7 | 0 | 5 | 500 | 300 | 300 | 0 | 500 | Lod following | 0.200 | 20,084,816 | 23.2 | 677.63 (10.4%) |
| 8 | 1000 | 0 | 0 | 300 | 500 | 280 | 500 | Generator order | 0.205 | 21,287,102 | 17.9 | 211.81 (2.2%) |

The amount of energy required for water heating per year is approximately 530 MWh, which can be fulfilled by the diversion load as the excess energy from the IHMS. The storage tank of 550 MWh/year has been implemented to use excess energy properly.

Wind turbines and PV boards are intended to be under a heap while working. For a breeze turbine, the heap is quite often an electrical burden, which is drawing power from the breeze turbine's generator. The two most basic burdens for a breeze turbine are (1) a battery bank and (2) an electrical lattice. Even though this is undoubtedly notable to a large number of you perusing this article, it is imperative to comprehend that an electrical burden (for example, battery bank or the electric lattice) keeps a breeze turbine in its structured working extent.

Too truly commute home this point, how about we consider as a relationship utilizing a hand drill on a bit of wood. For our relationship, the hand drill is the breeze turbine and the wood is the electrical burden. If the hand drill is gone to its most powerful setting and permitted to turn in free air, it will likely turn at around 700 rpm. This is the "no heap" circumstance because the drill is not accomplishing any work. Presently, if we utilize the hand drill on its most powerful setting to begin to penetrate a gap in the wood, what will occur? The rpm of the hand drill will hinder a ton contrasted with when it was turning in free air. This is because the drill presently needs to make a solid effort to make the gap in the wood. This is the "stacked" circumstance. Presently, a drill is intended to work under "no heap" however, a breeze turbine is not. If a breeze turbine or a PV module works under no heap in high wind conditions, it can fall to pieces. In high breezes and no heap, the breeze turbine edges can turn so quick that the cutting edges can come ripping off or, at any rate, put extreme anxieties and strains on the breeze turbine segments, which will make them destroy rapidly. Or on the other hand, at the end of the day, a breeze turbine works securely and appropriately when it is under a heap.

As expressed already, wind turbines are commonly used to charge battery banks or feed an electrical network. Both of these applications

**TABLE 6** Enforcement of structure components in diverse optimization results

| Position | Active hours of DG in a year | | | Percentage of individual contribution | | | | | $CO_2$ (kg/year) |
|---|---|---|---|---|---|---|---|---|---|
| | GEN 1 | GEN 2 | GEN 3 | PV | Wind | GEN 1 | GEN 2 | GEN 3 | |
| 1 | 3687 | 3977 | 3779 | 13.07 | 28.89 | 7.36 | 21.13 | 35.05 | 1,735,836 |
| 2 | 3891 | 4491 | 3877 | 0 | 6.32 | 21.19 | 33.16 | 39.34 | 2,963,826 |
| 3 | 0 | 6311 | 2228 | 15.31 | 32.35 | 0 | 46.69 | 7.64 | 2,726,352 |
| 4 | 4365 | 3905 | 3726 | 21.72 | 0 | 8.24 | 22.76 | 47.28 | 3.376,100 |
| 5 | 0 | 7269 | 1269 | 0 | 41.0 | 0 | 53.30 | 7.70 | 3,037,515 |
| 6 | 5442 | 6144 | 3442 | 9.0 | 28.22 | 39.36 | 19.22 | 6.20 | 3,168,826 |
| 7 | 5706 | 5619 | 2489 | 0 | 35.82 | 41.98 | 17.85 | 6.35 | 3,251,200 |
| 8 | 0 | 5047 | 6472 | 22.51 | 0 | 0 | 27.25 | 51.24 | 3,438,323 |
| 9 | 5680 | 6550 | 3759 | 16.45 | 0 | 50.86 | 25.28 | 8.42 | 3,705,215 |
| 10 | 4576 | 4931 | 6116 | 0 | 0 | 9.96 | 29.83 | 60.21 | 5,124,879 |
| 11 | 6773 | 5575 | 4169 | 0 | 0 | 64.87 | 24.81 | 10.32 | 4,215,986 |

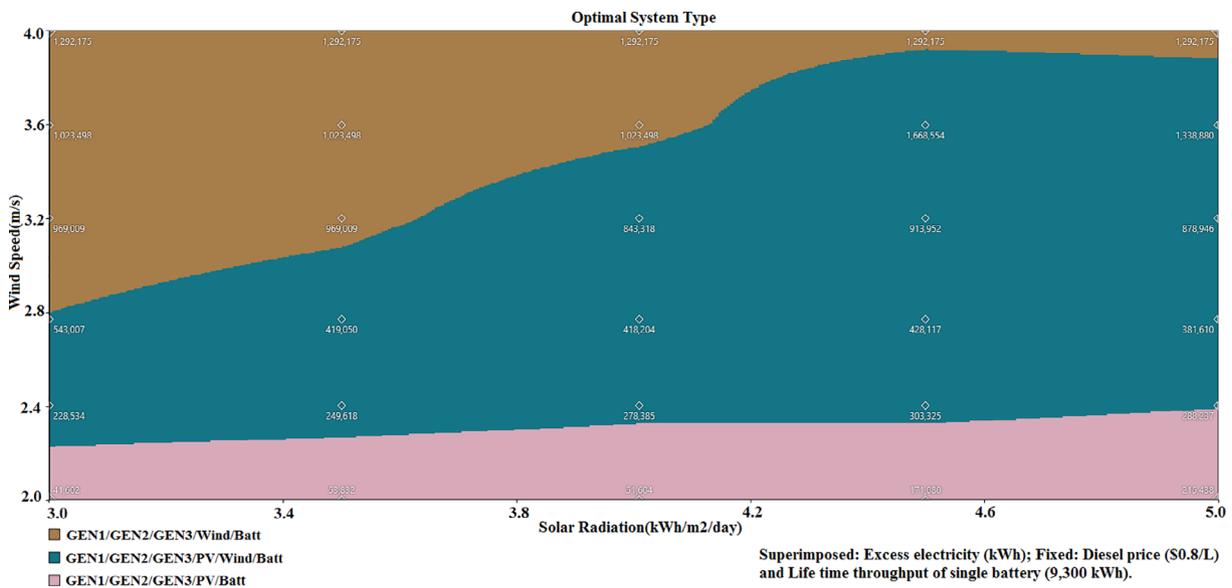

**FIGURE 7** System design with variable wind speed and solar radiation resulting from sensitivity analysis [Color figure can be viewed at wileyonlinelibrary.com]

required dump stacks, however, how about we analyze the battery bank application in more detail.

A breeze turbine will keep on charging a battery bank until completely energized. For a 12-volt battery bank, this is around 14 volts (The specific completely energized voltage of a 12-volt battery bank relies upon the kind of batteries being utilized). When the battery bank is completely energized, when wind turbine stops charging the battery storage there certain outages may occur for few moments (for example, battery annihilation, the danger of blast, and so on.) But, hold up there is an issue; We need to hold the breeze turbine under an electrical burden. To achieve this undertaking a redirection load charge controller is utilized.

In the most straightforward terms, a redirection load charge controller is a voltage sensor switch. The charge controller continually screens the voltage of the battery bank. On account of a 12-volt battery bank, when the voltage level arrives at around 14 volts, the charge controller detects this and separates the breeze turbine from the battery bank. Presently, we said that a preoccupation load charge controller is a voltage sensor switch. Thus, a redirection load charge controller does not fit for disengaging the breeze turbine from the battery bank, it is additionally fit for changing the breeze turbine's association with the preoccupation load. What is more, this is actually what the preoccupation load charge controller does, which holds the breeze turbine under a consistent electrical burden.

When the battery bank's voltage drops a little (roughly 13.6 volts for a 12-volt battery bank), the charge controller detects this and switches the breeze turbine back to charging the battery bank. This cycle is rehashed as fundamental, which keeps the battery bank from cheating and the breeze turbine consistently under the burden.

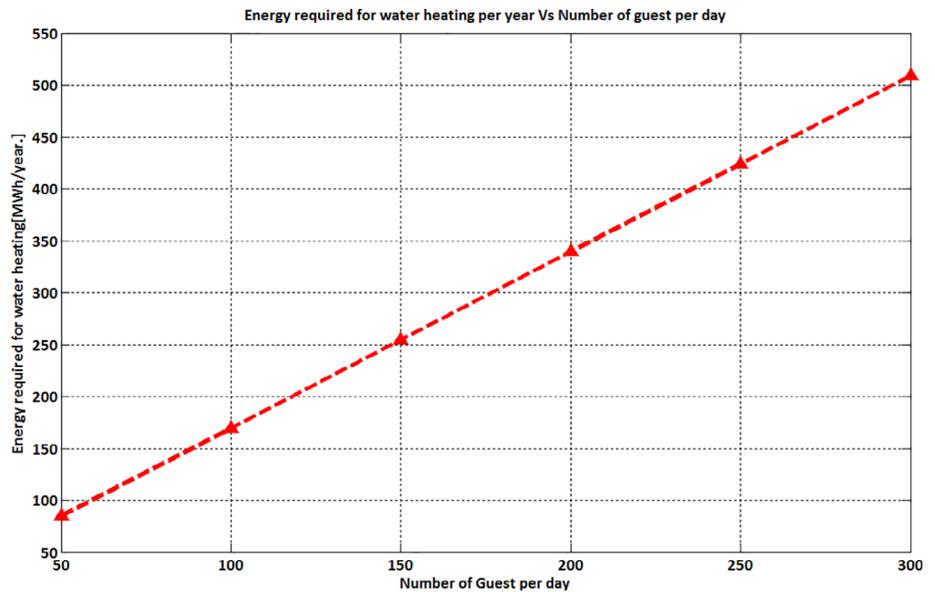

**FIGURE 8** Yearly energy required for water heating with a different number of guest arrival [Color figure can be viewed at wileyonlinelibrary.com]

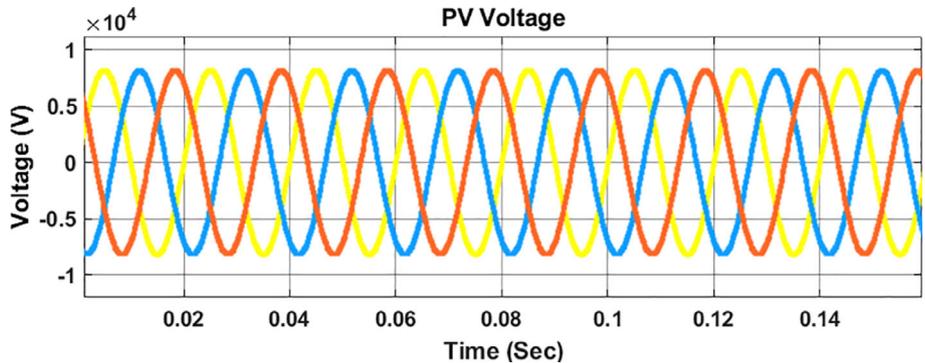

**FIGURE 9** PV voltage from the PV-Wind-DG-Battery combinedly as a hybrid alternative energy system [Color figure can be viewed at wileyonlinelibrary.com]

### 3.1 | Sensitivity analysis

Solar radiation and wind speed are the sensitivity constrain where the solar radiation was varied within 3.0–5.0 kWh/m²/day by a step of 0.4 and the mounted yearly typical solar irradiance for Penang Island has been determined to be 4.02 kWh/m²/day. However, the wind speed again differed from 2.0 to 4.0 m/s with a step size of 0.4 having a scaled 12-month average wind speed of 2.77 m/s. On Figure 7, it has been detected that, whenever wind speed is greater than 2.25 m/s, as well as solar radiation, is superior to 3.5 kWh/m²/day, wind/diesel/PV/battery stands to be the best combination to supply electricity throughout the resort. Subsequently, if the wind speediness is larger than 3.25 m/s and solar radiation is lower as 3.5 kWh/m²/day then wind/diesel/battery becomes the best combination to supply electricity within the resort. Figure 7 shows the System design with variable wind speed and solar radiation resulting from sensitivity analysis.

### 3.2 | Effective utilization of excess renewable energy using an electrical heater provided with the storage tank as a diversion load

A case study of water consumption for two Malaysian resort was conducted by F.E. Tang.[15] This work indicates that water consumption is 500 L per guest per day in the Malaysian resort. As mentioned in Section 2.2, the resort has 312 rooms and suites, two restaurants, four meeting rooms, one ballroom, and so on. In addition, the rooms of deluxe and superior chalets have bathtub facilities. Daily (monthly or yearly) water consumption and guest arrival data were not available at the resort. As mentioned in Table 5, the best optimized IHMS for Penang Hill resort has 517.29 MWh of surplus energy per year. In this study, a detailed analysis has been done by assuming water consumption of 200 L per guest per day with the variable guest arrival for proper utilization of this excess energy per annum as shown in Figure 8. The energy required to heat water was calculated by using the following formula[13]:

$$E_h = M_w \times S_w \times (T_f - T_{in}) \quad (13)$$

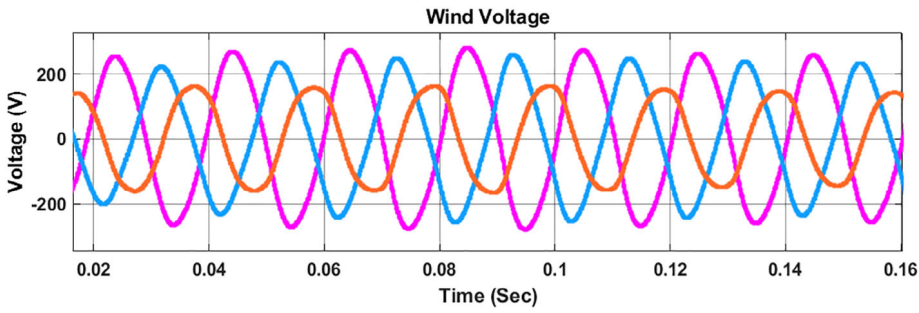

**FIGURE 10** Wind voltage from the PV-Wind-DG-Battery combinedly as a hybrid alternative energy system [Color figure can be viewed at wileyonlinelibrary.com]

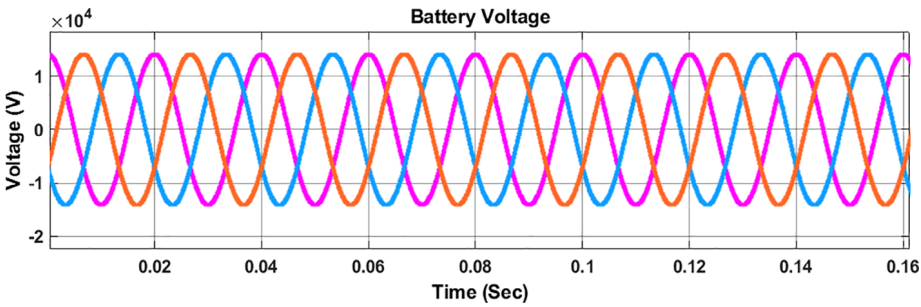

**FIGURE 11** Battery voltage from the PV-Wind-DG-Battery combinedly as a hybrid alternative energy system [Color figure can be viewed at wileyonlinelibrary.com]

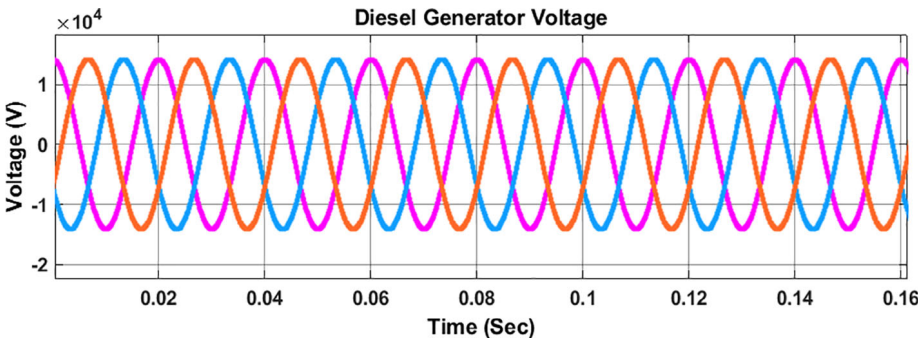

**FIGURE 12** Diesel Generator voltage from the PV-Wind-DG-Battery combinedly as a hybrid alternative energy system [Color figure can be viewed at wileyonlinelibrary.com]

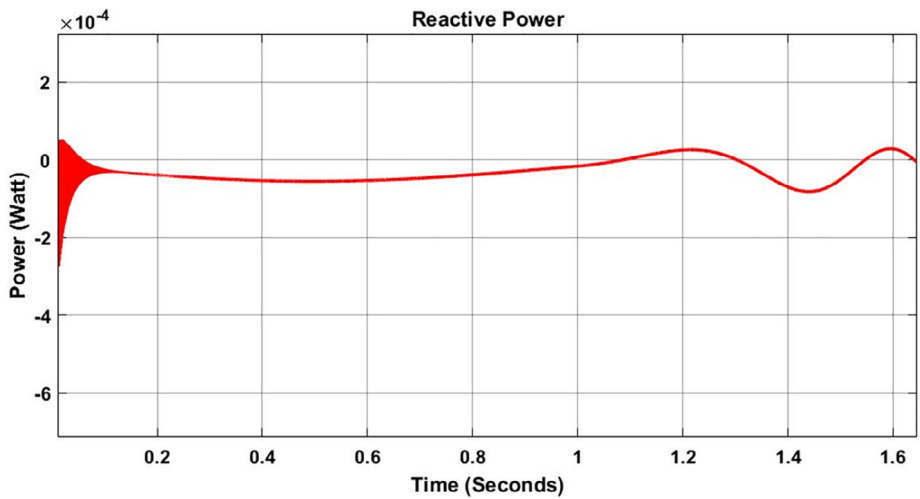

**FIGURE 13** Power production as the reactive power from the PV-Wind-DG-Battery combinedly as a hybrid alternative energy system [Color figure can be viewed at wileyonlinelibrary.com]

where $M_w$ is mass of water in kg, $S_w$ is the specific heat of the water in KJ/kg/°C, which is 4.18 KJ/kg/°C, $T_f$ is final temperature, and $T_{in}$ is the initial temperature. Figure 8 shows that the excess energy of 413.19 MWh is effectively utilized with the guest arrival of 250/day.

The results of using proportional–integral–derivative (PID) controller with automatic tuning are shown in the next few figures. Figure 9 shows the voltage of the PV module of the designed IHMS. The voltage is held in between $-1 \times 10^4$ and $1 \times 10^4$ V. According to the optimal size of the PV module, it created some imbalance in the three-phase voltage. The PV voltage stabilized by the three-phase programmable voltage source and tuned PID controller. Yellow line denotes the phase-1, the blue line indicates the phase-2, and red line denotes the phase-3. Figure 10 shows the tuned and controlled voltage of wind module of the IHMS in-between −200 and 200 V. Yellow line denotes the phase-1, the pink line indicates the phase-2, and blue line denotes the phase-3. The wind voltage stabilized by the tuned PID controller and active power acceleration by the diesel generator. Figure 11 shows the tuned and controlled voltage of the battery module of the IHMS in-between $-1 \times 10^4$ and $1 \times 10^4$ V. Pink line denotes the phase-1, the red line indicates the phase-2, and blue line denotes the phase-3. Battery voltage tuned and stabilized by the reactive power support from the diesel generator. Figure 12 shows the tuned and controlled voltage of diesel generator module of the IHMS in-between $-1 \times 10^4$ and $1 \times 10^4$ V. Pink line denotes the phase-1, the red line indicates the phase-2, and blue line denotes the phase-3. Diesel generator voltage keeps the system voltage stable with the dedicated frequency for the distribution network. Figure 13 shows the reactive power flow for 0–1.6 s and well-tuned. Reactive power response has the deviation toward negative from the beginning to until 0.6 s. Then it started rising and at 1.12 s it was going down again. After 1.14 s it was again rising and stayed stable until 1.6 s. The reactive power response variations occurred due to the stochastic behavior of renewable resources as well as the system loses in power electronics devices. Figure 14 shows the active power flow with tuned by PID control technique for 0 to 1.4 s. Active power response has the deviation toward positive era from the beginning to until 1 s. Then it started falling and at 1.12 s it was going down again. After 1.25 s it was again rising and stayed stable until 1.6 s.

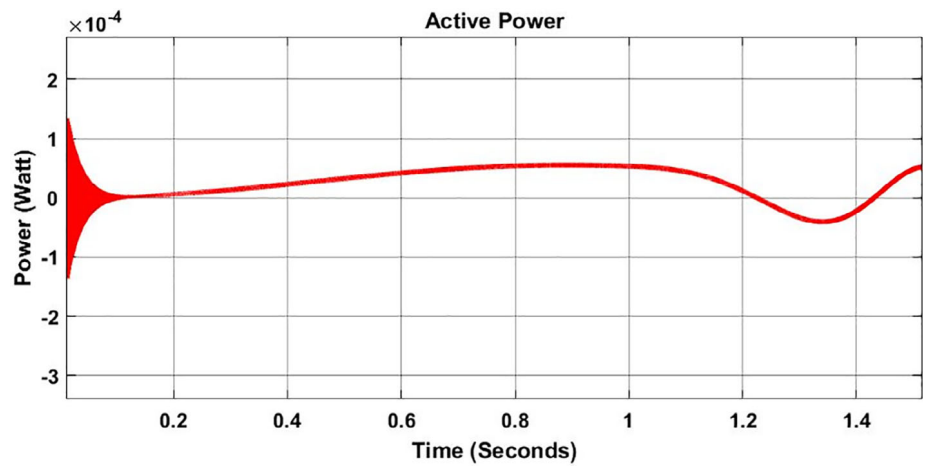

**FIGURE 14** Power production as the active power from the PV-Wind-DG-Battery combinedly as a hybrid alternative energy system [Color figure can be viewed at wileyonlinelibrary.com]

## 4 | CONCLUSION

The remote and decentralized tourism sectors in Malaysia are being dependent on fossil fuels as diesel generators and other conventional power sources for their Electrical power loads. The diesel-based power system emits a large amount of $CO_2$ and GHG, which are considered to be very harmful to the tourist zone. Since the cost of diesel fuel gets higher in remote islands compared to the mainland, therefore, a schematic comprising of IHMS may be an environment-friendly solution in providing electricity to the resort along with being cost-efficient. However, excess electricity generated from this hybrid RE system can be considered dangerous for the batteries due to the possibility of overcharging. Hence, in this article, an optimum amalgamation of an IHMS has been proposed for the Penang Hill Resort, located at Penang Island Malaysia, with effective utilization of excess energy through diversion load. In addition to this, a power management strategy for the diversion load has also been developed and discussed where the utilization of excess electricity is analyzed in terms of water consumption and tourist arrival in the resort. From this study, the prime-optimized scheme was contrived of a 600 kW PV, 3 units of diesel generator, four wind turbine, a bidirectional converter of 500 kW, and 200 batteries. Concerning this, the nominated wind turbine for the location has been determined in having a low cut-in wind speediness of 3–3.5 m/s. Along with this, on a wind swiftness of 7–8 m/s, the turbine gives a rated output of 250 kW. In this conducted study, the NPC, COE, and RF of the enhanced structure stayed as $0.165/kWh, $20.66 m, 31.5%, respectively, where the amount of surplus energy was 517.29 MWh/year. According to the result, the aforementioned amount of excess energy can be properly utilized using an electrical heater as a diversion load. The diesel generator is the only system for the resort, which emits almost twice the GHG than the best-optimized hybrid system. The future work for this project will be to enhance the battery efficiency and the reliability of the green energy supply. Transmission connection from the mainland is not required, as electricity can be generated from renewable energy sources and it will decentralize electricity supply. Therefore, as the weather is not constant every day, the production of electricity will not be the same according to the demand for the load. The limitation of this project is that the instrument's setup cost and maintenance cost is a bit expensive. However, a backup generator should be attached due to an adverse circumstance to reduce the chance of blackout.

## ACKNOWLEDGMENT
This research work has been completed by the RITFS Scholarship of RMIT University, Melbourne, Australia and the RITFS research grant funded by the Federal Government of Australia.

## AUTHOR CONTRIBUTIONS

Sk. A. Shezan: Conceptualization; supervision; data curation; formal analysis; methodology; software; writing-original draft; Project administration; Funding acquisition; Resources. Rawdah Rawdah: Conceptualization; data curation; formal analysis; writing-original draft; Validation. S. Shafin Ali: Formal analysis; Software; writing-original draft; Investigation; Validation. Ziaur Rahman: Software; methodology; writing-review and editing; Visualization.

## DATA AVAILABILITY STATEMENT
Figure 2, 3, and 4 describes the data used for this analysis.

## ORCID
*SK. A. Shezan*  https://orcid.org/0000-0003-3636-8977
*S. Shafin Ali*  https://orcid.org/0000-0002-9626-5411
*Ziaur Rahman*  https://orcid.org/0000-0002-7759-3428

## REFERENCES
1. Uddin M, Mukherjee S, Chang H, Lakshman TV. Sdn-based multi-protocol edge switching for IoT service automation. *IEEE J Select Areas Commun*. 2018;36(12):2775-2786.
2. Taylor D, Turner S, Willette D, Uawithya P. De-centralized Electricity in Africa and Southeast Asia: Issues and Solutions. A Study by Accenture Development Partnerships. https://assets.rockefellerfoundation.org/app/uploads/20150128205930/De-centralized-Electricity-in-Africa-and-Southeast-Asia.pdf. Accessed July 2, 2020.
3. Basir Khan MR, Jidin R, Pasupuleti J. Optimal combination of solar, wind, micro-hydro and diesel systems based on actual seasonal load profiles for a resort Island in the South China Sea. *Energy*. 2015;82:80-97.
4. Shezan SA, Ullah KR, Hossain A, Chong WT, Julai S. Feasibility analysis of a hybrid off-grid wind–DG-battery energy system for the eco-tourism remote areas. *Clean Technol Environ Policy*. 2015;17(8):2417-2430.
5. Ashourian MHA, Cheratia SM, Mohd A, Niknama ZN, Mokhtara AS, Anwarib M. Optimal green energy management for Island resorts in Malaysia. *Renew Energy*. 2013;51:36-45.
6. Fadaeenejad M, Radzi MAM, AbKadir MZA, Hizam H. Assessment of hybrid renewable power sources for rural electrification in Malaysia. *Renew Sustain Energy Rev*. 2014;30:299-305.
7. Anwari M. An evaluation of hybrid wind/diesel energy potential in Pemanggil Island Malaysia. Paper presented at: Power Engineering and Renewable Energy (ICPERE), 2012 International Conference on 2012.
8. Tobe J. Managing Battery Charging Using Diversion Loads; March/April 2015.
9. Jon R. Luoma:The challenge for green energy: How to store excess electricity; 13 July, 2009.
10. Esan AB, Agbetuyi AB, Oghorada AF, et al. Reliability assessments of an islanded hybrid PV-diesel-battery system for a typical rural community in Nigeria. *Heliyon*. 2019;5(5):e01632.
11. Singh A, Baredar P, Gupta B. Computational simulation & optimization of a solar, fuel cell and biomass hybrid energy system using HOMER pro software. *Procedia Eng*. 2015;127:743-750.
12. Singh A, Baredar P. Techno-economic assessment of a solar PV, fuel cell, and biomass gasifier hybrid energy system. *Energy Rep*. 2016;2:254-260.
13. Ismail MS. Effective utilization of excess energy in standalone hybrid renewable energy systems for improving comfort ability and reducing cost of energy: a review and analysis. *Renew Sustain Energy Rev*. 2015;42:726-734.
14. Zahir E, Akhtar Shihab S, Akash SM, Shafin Ali S, Mabruba N. Introducing a voice synthesized visual and auditory navigator implemented for a University Campus. Paper presented at: 2018 IEEE International WIE Conference on Electrical and Computer Engineering (WIECON-ECE), Chonburi, Thailand, 2018, pp. 133–136. https://doi.org/10.1109/WIECON-ECE.2018.8783060.
15. Rashid S. Optimized design of a hybrid PV-wind-diesel energy system for sustainable development at coastal areas in Bangladesh. *Environ Prog Sustain Energy*. 2017;36(1):297-304.
16. Demiroren A, Yilmaz U. Analysis of change in electric energy cost with using renewable energy sources in Gökceada, Turkey: an Island example. *Renew Sustain Energy Rev*. 2010;14(1):323-333.
17. Ismail MS, Moghavvemi M, Mahlia TMI. Techno-economic analysis of an optimized photovoltaic and diesel generator hybrid power system for remote houses in a tropical climate. *Energ Conver Manage*. 2013;69:163-173.
18. Dalton GJ, Lockington DA, Baldock TE. Feasibility analysis of stand-alone renewable energy supply options for a large hotel. *Renew Energy*. 2008;33(7):1475-1490.
19. Dalton GJ, Lockington DA, Baldock TE. Feasibility analysis of renewable energy supply options for a grid-connected large hotel. *Renew Energy*. 2009;34(4):955-964.
20. Shaahid SM, Al-Hadhrami LM, Rahman MK. Review of economic assessment of hybrid photovoltaic-diesel-battery power systems for residential loads for different provinces of Saudi Arabia. *Renew Sustain Energy Rev*. 2014;31:174-181.
21. Shezan SA, Khan N, Anowar MT, et al. Fuzzy logic implementation with MATLAB for solar-wind-battery-diesel hybrid energy system. *Imp J Interdiscip Res*. 2016;2(5):574-583.
22. Hafez O, Bhattacharya K. Optimal planning and design of a renewable energy based supply system for microgrids. *Renew Energy*. 2012;45:7-15.
23. Shezan SKA, Farzana M, Hossain A, Ishrak A. Techno-economic and feasibility analysis of a micro-grid wind-dg-battery hybrid energy system for remote and decentralized areas. *Int J Adv Eng Technol (IJAET)*. 2015;8(6):874-888.
24. Muda A, Omar CMC, Ponrahono Z, Shamsuddin K, Chung D, Gambaris ATK. Tioman as international tourism Island: in perspective of planning development, management and guidelines. *Contemporary Environmental Quality Management in Malaysia and Selected Countries*. UPM Press, Universiti Putra Malaysia, 43400 Serdang, Selangor, Malaysia; 2011.
25. Talukder M, Rawdah, Aktar A, Neelima A, Rahman A. EOG based home automation system by cursor movement using a graphical user interface (GUI). Paper presented at: 2018 IEEE International WIE Conference on Electrical and Computer Engineering (WIECON-ECE), Chonburi, Thailand, 2018, pp. 1–4. https://doi.org/10.1109/WIECON-ECE.2018.8783025.
26. Hossen MD, Shezan SA. Optimization and assessment of a hybrid solar-wind-biomass renewable energy system for Kiribati Island. *Int J Res Eng Sci*. 2019;6(10):1-8.
27. Shezan SKA, Julai S, Kibria MA, et al. Performance analysis of an off-grid wind-PV (photovoltaic)-diesel-battery hybrid energy system feasible for remote areas. *J Clean Prod*. 2016;125:121-132.
28. Ohunakin O, Adaramola M, Oyewola O. Wind energy evaluation for electricity generation using WECS in seven selected locations in Nigeria. *Appl Energy*. 2011;88(9):3197-3206.

29. Shezan S, Lai CY. Optimization of hybrid wind-diesel-battery energy system for remote areas of Malaysia. Paper presented at: 2017 Australasian Universities Power Engineering Conference (AUPEC). 2017. IEEE.
30. Akdağ S, Bagiorgas H, Mihalakakou G. Use of two-component Weibull mixtures in the analysis of wind speed in the eastern Mediterranean. *Appl Energy*. 2010;87(8):2566-2573.
31. Islam MR, Saidur R, Rahim NA. Assessment of wind energy potentiality at Kudat and Labuan, Malaysia using Weibull distribution function. *Energy*. 2011;36(2):985-992.
32. Arefin SS, Das N. Optimized hybrid wind-diesel energy system with feasibility analysis. *Technol Econ Smart Grids Sustain Energy*. 2017;2(1):9.
33. Shezan SKA, Al-Mamoon HWP. Performance investigation of an advanced hybrid renewable energy system in Indonesia. *Environ Prog Sustain Energy*. 2018;37(4):1424-1432.
34. Shezan SA. Optimization and assessment of an off-grid photovoltaic–diesel–battery hybrid sustainable energy system for remote residential applications. *Environ Prog Sustain Energy*. 2019;38(6):e13340.